\documentclass[twocolumn,superscriptaddress,showpacs,preprintnumbers,amsmath,amssymb,prl]{revtex4}
\usepackage{graphicx}
\usepackage{dcolumn}
\usepackage{bm}

\def\ra{\rightarrow}
\newcommand{\GeV}{\ensuremath{\mathrm{Ge\kern -0.1em V}}}
\newcommand{\MeV}{\ensuremath{\mathrm{Me\kern -0.1em V}}}
\newcommand{\cm}{\ensuremath{\mathrm{cm}}}
\newcommand{\um}{\ensuremath{\mathrm{\mu m}}}
\newcommand{\pp}{\ensuremath{p\overline{p}}}
\newcommand{\pb}{\ensuremath{\mathrm{pb}^{-1}}}
\newcommand{\bs}{\ensuremath{B_s^0}}
\newcommand{\bd}{\ensuremath{B_d^0}}
\newcommand{\bsd}{\ensuremath{B_{s,d}^0}}
\newcommand{\bu}{\ensuremath{B^{+}}}
\newcommand{\mm}{\ensuremath{\mu^{+}\mu^{-}}}

\newcommand{\hh}{\ensuremath{h^{+}h^{-}}}
\newcommand{\bsmm}{\ensuremath{\bs\ra\mm}}
\newcommand{\bdmm}{\ensuremath{\bd\ra\mm}}
\newcommand{\bsdmm}{\ensuremath{\bsd\ra\mm}}
\newcommand{\bjk}{\ensuremath{\bu\ra J/\psi K^{+}}}
\newcommand{\brbsmm}{\ensuremath{\mathcal{B}(\bsmm)}}
\newcommand{\brbdmm}{\ensuremath{\mathcal{B}(\bdmm)}}
\newcommand{\brbsdmm}{\ensuremath{\mathcal{B}({B_{s,d}^0}\ra\mm)}}
\newcommand{\Mmm}{\ensuremath{M_{\mu\mu}}}

\newcommand{\Lxyz}{\ensuremath{{L}_{\rm{3D}}}}
\newcommand{\ctau}{\ensuremath{\lambda}}
\newcommand{\pting}{\ensuremath{\Delta\Theta}}
\newcommand{\iso}{\ensuremath{\mathit{I}}}
\newcommand{\ptmm}{\ensuremath{\vec{p}^{\:\mu\mu}_{T}}}
\newcommand{\pmm}{\ensuremath{\vec{p}^{\:\mu\mu}}}
\newcommand{\cdf}{CDF~II}

\begin{document}

\title{{\bf{ Search for \bsmm\ and \bdmm\ Decays 
        in $p\overline{p}$ Collisions  with CDF II}} }

\affiliation{Institute of Physics, Academia Sinica, Taipei, Taiwan 11529, Republic of China} 
\affiliation{Argonne National Laboratory, Argonne, Illinois 60439} 
\affiliation{Institut de Fisica d'Altes Energies, Universitat Autonoma de Barcelona, E-08193, Bellaterra (Barcelona), Spain} 
\affiliation{Baylor University, Waco, Texas  76798} 
\affiliation{Istituto Nazionale di Fisica Nucleare, University of Bologna, I-40127 Bologna, Italy} 
\affiliation{Brandeis University, Waltham, Massachusetts 02254} 
\affiliation{University of California, Davis, Davis, California  95616} 
\affiliation{University of California, Los Angeles, Los Angeles, California  90024} 
\affiliation{University of California, San Diego, La Jolla, California  92093} 
\affiliation{University of California, Santa Barbara, Santa Barbara, California 93106} 
\affiliation{Instituto de Fisica de Cantabria, CSIC-University of Cantabria, 39005 Santander, Spain} 
\affiliation{Carnegie Mellon University, Pittsburgh, PA  15213} 
\affiliation{Enrico Fermi Institute, University of Chicago, Chicago, Illinois 60637} 
\affiliation{Joint Institute for Nuclear Research, RU-141980 Dubna, Russia} 
\affiliation{Duke University, Durham, North Carolina  27708} 
\affiliation{Fermi National Accelerator Laboratory, Batavia, Illinois 60510} 
\affiliation{University of Florida, Gainesville, Florida  32611} 
\affiliation{Laboratori Nazionali di Frascati, Istituto Nazionale di Fisica Nucleare, I-00044 Frascati, Italy} 
\affiliation{University of Geneva, CH-1211 Geneva 4, Switzerland} 
\affiliation{Glasgow University, Glasgow G12 8QQ, United Kingdom} 
\affiliation{Harvard University, Cambridge, Massachusetts 02138} 
\affiliation{Division of High Energy Physics, Department of Physics, University of Helsinki and Helsinki Institute of Physics, FIN-00014, Helsinki, Finland} 
\affiliation{University of Illinois, Urbana, Illinois 61801} 
\affiliation{The Johns Hopkins University, Baltimore, Maryland 21218} 
\affiliation{Institut f\"{u}r Experimentelle Kernphysik, Universit\"{a}t Karlsruhe, 76128 Karlsruhe, Germany} 
\affiliation{High Energy Accelerator Research Organization (KEK), Tsukuba, Ibaraki 305, Japan} 
\affiliation{Center for High Energy Physics: Kyungpook National University, Taegu 702-701; Seoul National University, Seoul 151-742; and SungKyunKwan University, Suwon 440-746; Korea} 
\affiliation{Ernest Orlando Lawrence Berkeley National Laboratory, Berkeley, California 94720} 
\affiliation{University of Liverpool, Liverpool L69 7ZE, United Kingdom} 
\affiliation{University College London, London WC1E 6BT, United Kingdom} 
\affiliation{Massachusetts Institute of Technology, Cambridge, Massachusetts  02139} 
\affiliation{Institute of Particle Physics: McGill University, Montr\'{e}al, Canada H3A~2T8; and University of Toronto, Toronto, Canada M5S~1A7} 
\affiliation{University of Michigan, Ann Arbor, Michigan 48109} 
\affiliation{Michigan State University, East Lansing, Michigan  48824} 
\affiliation{Institution for Theoretical and Experimental Physics, ITEP, Moscow 117259, Russia} 
\affiliation{University of New Mexico, Albuquerque, New Mexico 87131} 
\affiliation{Northwestern University, Evanston, Illinois  60208} 
\affiliation{The Ohio State University, Columbus, Ohio  43210} 
\affiliation{Okayama University, Okayama 700-8530, Japan} 
\affiliation{Osaka City University, Osaka 588, Japan} 
\affiliation{University of Oxford, Oxford OX1 3RH, United Kingdom} 
\affiliation{University of Padova, Istituto Nazionale di Fisica Nucleare, Sezione di Padova-Trento, I-35131 Padova, Italy} 
\affiliation{University of Pennsylvania, Philadelphia, Pennsylvania 19104} 
\affiliation{Istituto Nazionale di Fisica Nucleare Pisa, Universities of Pisa, Siena and Scuola Normale Superiore, I-56127 Pisa, Italy} 
\affiliation{University of Pittsburgh, Pittsburgh, Pennsylvania 15260} 
\affiliation{Purdue University, West Lafayette, Indiana 47907} 
\affiliation{University of Rochester, Rochester, New York 14627} 
\affiliation{The Rockefeller University, New York, New York 10021} 
\affiliation{Istituto Nazionale di Fisica Nucleare, Sezione di Roma 1, University of Rome ``La Sapienza," I-00185 Roma, Italy} 
\affiliation{Rutgers University, Piscataway, New Jersey 08855} 
\affiliation{Texas A\&M University, College Station, Texas 77843} 
\affiliation{Istituto Nazionale di Fisica Nucleare, University of Trieste/\ Udine, Italy} 
\affiliation{University of Tsukuba, Tsukuba, Ibaraki 305, Japan} 
\affiliation{Tufts University, Medford, Massachusetts 02155} 
\affiliation{Waseda University, Tokyo 169, Japan} 
\affiliation{Wayne State University, Detroit, Michigan  48201} 
\affiliation{University of Wisconsin, Madison, Wisconsin 53706} 
\affiliation{Yale University, New Haven, Connecticut 06520} 
\author{A.~Abulencia}
\affiliation{University of Illinois, Urbana, Illinois 61801}
\author{D.~Acosta}
\affiliation{University of Florida, Gainesville, Florida  32611}
\author{J.~Adelman}
\affiliation{Enrico Fermi Institute, University of Chicago, Chicago, Illinois 60637}
\author{T.~Affolder}
\affiliation{University of California, Santa Barbara, Santa Barbara, California 93106}
\author{T.~Akimoto}
\affiliation{University of Tsukuba, Tsukuba, Ibaraki 305, Japan}
\author{M.G.~Albrow}
\affiliation{Fermi National Accelerator Laboratory, Batavia, Illinois 60510}
\author{D.~Ambrose}
\affiliation{Fermi National Accelerator Laboratory, Batavia, Illinois 60510}
\author{S.~Amerio}
\affiliation{University of Padova, Istituto Nazionale di Fisica Nucleare, Sezione di Padova-Trento, I-35131 Padova, Italy}
\author{D.~Amidei}
\affiliation{University of Michigan, Ann Arbor, Michigan 48109}
\author{A.~Anastassov}
\affiliation{Rutgers University, Piscataway, New Jersey 08855}
% ...
\collaboration{CDF Collaboration}
\noaffiliation

\begin{abstract}
  We report on a search for \bsmm\ and \bdmm\ decays in \pp\ collisions
  at $\sqrt{s} = 1.96$~TeV using $364\:\pb$ of data collected by the \cdf\ 
  detector at the Fermilab Tevatron Collider.
  After applying all selection requirements, we observe no candidates inside the 
  \bs\ or \bd\ mass windows.  The 
  resulting upper limits on the branching fractions are 
  $\brbsmm < 1.5\times10^{-7}$ and $\brbdmm < 3.9\times10^{-8}$ at 
  $90\%$ confidence level.
\end{abstract}


\maketitle


In the standard model (SM), Flavor Changing Neutral Current (FCNC) decays are 
highly suppressed and can only occur through higher order diagrams. 
The decay rate for the FCNC decay \bsmm~\cite{qconj} is proportional to the
CKM matrix element $\left|V_{ts}\right|^{2}$. The rate of \bdmm\ decays is further 
suppressed by the ratio of $\left|V_{td}/V_{ts}\right|^2$.  
The SM expectations for these branching fractions are 
$\brbsmm=(3.42\pm0.54)\times10^{-9}$ and 
$\brbdmm=(1.00\pm0.14)\times10^{-10}$~\cite{smbr}, which are about two orders 
of magnitude smaller than the current experimental sensitivity.  
However, new physics contributions can significantly enhance these branching 
fractions.  An observation of these decays at the Tevatron would be unambiguous 
evidence for physics beyond the SM.  The best existing experimental 
bound on ${\mathcal{B}}$ is $<4.1\times10^{-7}$~\cite{D0prl} 
($<8.3\times 10^{-8}$~\cite{BABARprl}) for $\bs (\bd) \rightarrow \mu^+\mu^-$
at the 90\% confidence level (C.L.).

In minimal supersymmetric (SUSY) extensions of the SM, additional diagrams 
involving SUSY particles also contribute to FCNC decay rates and the branching
fraction $\brbsdmm\propto(\tan\beta)^{6}$, where $\tan\beta$ is the ratio of 
vacuum expectation values of the two neutral CP-even Higgs fields.  Large values of 
$\tan\beta$ enhance 
the decay rate to a level observable by the Tevatron experiments~\cite{tanb}.  
For example, increases of one to three orders of magnitude are obtained in the 
minimal SO(10) models~\cite{SO10}, which favor large values of $\tan\beta$. 
For the minimal flavor violating (MFV) models, \bdmm\ 
remains suppressed relative to \bsmm\ due to  
$\left|V_{td}/V_{ts}\right|^2$.  This may not be true for non-MFV 
models such as R-parity violating SUSY~\cite{RPV}, which can produce large 
enhancements, even for low values of $\tan\beta$, in either or both of the $B_s$ 
and $B_d$ FCNC decay rates. 
Thus, a simultaneous observation of \bsdmm\ decays can be important in 
determining the flavor structure of the new physics. In the absence of 
an observation, any improvements to the limits can be used to constrain 
significantly many SUSY models~\cite{tanb,SO10,RPV,Models}.

In this Letter, we report on a search for \bsmm\ and \bdmm\ decays using 
364~\pb~of data collected by the upgraded Collider Detector at Fermilab (\cdf). 
This data set includes 171~\pb~of data from our previous measurement~\cite{PRL1}. 
We significantly improve the sensitivity of the search over our former 
analysis by doubling our data sample, extending the muon 
acceptance, and using a likelihood ratio technique for signal and 
background separation. The limits we present here are the most stringent to date
and supersede our previous results.

The \cdf\ detector is described in detail in Ref.~\cite{Acosta:2004yw}. 
The inner tracking system is composed of a silicon microstrip detector 
(SVX II)~\cite{SVX} surrounded by an open-cell wire drift chamber (COT)~\cite{COT}.  
These tracking detectors are immersed in a $1.4$~T magnetic field and measure $p_T$,
charged particle momentum in the plane transverse to the beamline.  Four layers 
of planar drift chambers (CMU)~\cite{CMU} detect muon candidates with $p_T>1.4~\GeV/c$ 
and provide coverage in the pseudorapidity range $\left|\eta\right|<0.6$, where 
$\eta = -\ln(\tan\frac{\theta}{2})$ and $\theta$ is the angle of the track with 
respect to the beamline.  The central muon extension (CMX) consists of conical 
sections of drift tubes and extends the coverage to $0.6<\left|\eta\right|<1.0$ 
for muon candidates with $p_T>2.0~\GeV/c$.  

The data used in this analysis are selected by two classes of dimuon 
triggers: for the CMU-CMU (U-U) triggers both muon candidates are reconstructed 
in the CMU chambers, while for the CMU-CMX (U-X) triggers one of the muon 
candidates is reconstructed in the CMX chambers.  The inclusion of the U-X 
trigger increases the signal acceptance by about 50\%.  The details of the trigger 
system and selection requirements can be found in Refs.~\cite{PRL1,Acosta:2004yw}.  
Since they have different sensitivities, we treat U-U and U-X channels separately, 
combining the results at the end.

The offline reconstruction begins by identifying two muon candidates of opposite charge 
which satisfy the online dimuon trigger requirements.  To avoid regions of rapidly 
changing trigger efficiency, we omit CMU (CMX) muon candidates with 
$p_{T}<2\ (2.2) \;\GeV/c$.  
The random combinatoric backgrounds are suppressed by requiring the vector sum 
of the muon transverse momenta to be $| \ptmm | > 4\;\GeV/c$.
The remaining pairs of muon tracks are then refit under the constraint that they
come from the same three-dimensional (3D) space point, and are required to 
satisfy vertex fit quality criteria.
The 3D decay length is given by $\Lxyz=\vec{L}\cdot\pmm/|\pmm|$, 
where $\vec{L}$ is the displacement vector from the primary to the dimuon vertex.  
The primary vertex is determined using a constrained vertex fit of all  
tracks in the event, excluding the \mm\ pair and other secondary decay tracks.  
For each $B$-candidate we 
estimate the proper decay time $\tau=\Mmm\Lxyz/|\pmm|$, where
\Mmm\ is the invariant mass and \pmm\ is the momentum vector of the dimuon system.
Additional background is reduced by demanding $\Lxyz < 1.0~\cm$ (to
remove poorly reconstructed tracks), the 
uncertainty on $\Lxyz$ to be less than 150 \um, and  
$2\sigma_{\ctau} < \ctau <0.3~\cm$, where $\ctau=c\cdot \tau$ and $\sigma_{\ctau}$ 
is the total uncertainty on $\ctau$.  There are 22459 (14305) dimuon candidates 
that fulfill all the above trigger and offline reconstruction requirements in the 
U-U (U-X) channel.  At this stage, the data sample is dominated by 
random combinatoric background.

We model the signal \bsdmm\ decays using the {\sc pythia} Monte Carlo (MC) 
program~\cite{pythia}. The {\sc pythia} events are passed through a full 
detector simulation and satisfy the same requirements as data. The $p_{T}$ 
spectrum of the $B$-hadron in the MC is weighted to match the measurement from 
Ref.~\cite{Acosta:2004yw}.  The signal MC samples are used primarily for analysis 
optimization and for estimating the efficiency of selection requirements.

\begin{figure}
  \includegraphics[width=8.0cm]{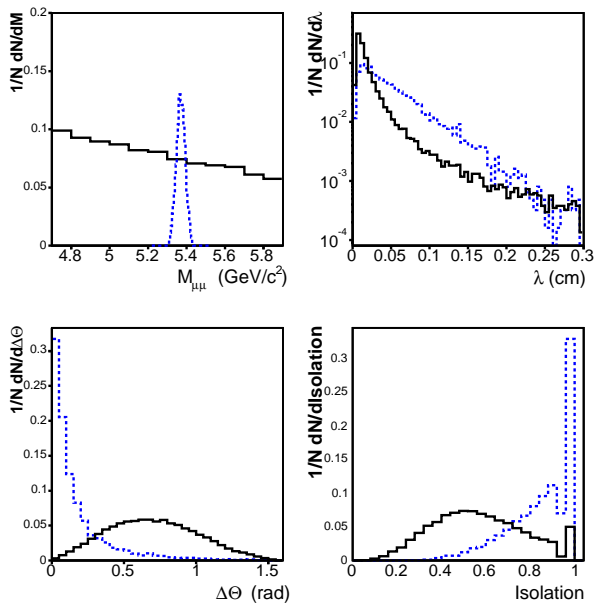}
  \caption{\label{fig:SvB} Distributions of the discriminating variables 
  for data (dominated by background) shown in solid histogram and signal 
  \bsmm\ events shown in dashed histogram.}
\end{figure}

For the final event selection we use the following four discriminating variables:
\Mmm, \ctau, the 3D opening angle \pting\ between vectors \pmm\ and $\vec{L}$,  
and the {\it B}-candidate track isolation (\iso)~\cite{iso}.  
We exploit long lifetime of {\it B} mesons to reject prompt combinatoric background
with the decay length (\ctau) requirement.  Additional combinatoric background and partially
reconstructed {\it B} hadrons are removed with the pointing angle requirement.  
Since {\it b}-quark fragmentation is hard, {\it B} hadrons carry most of the transverse 
momentum of the {\it b} quark, and thus are isolated.  We use the variable \iso\ to
enhance the heavy flavor content of the sample and also to reject partially
reconstructed {\it B} hadrons, which are less isolated.
Figure~\ref{fig:SvB} compares the distributions of these variables for 
data (which is background-dominated) to MC signal events.
Based on the observed distributions, we apply two additional loose requirements, 
$\iso>0.5$ and $\pting<0.7~{\rm{rad}}$, to further reject background while 
maintaining 92\% of signal efficiency.  We collectively refer to all
the selection requirements applied up to this point as the ``baseline''
requirements. In the data, 6242 (4908) events survive 
the baseline requirements in the U-U (U-X) sample.  

\begin{figure}
  \includegraphics[width=8.0cm]{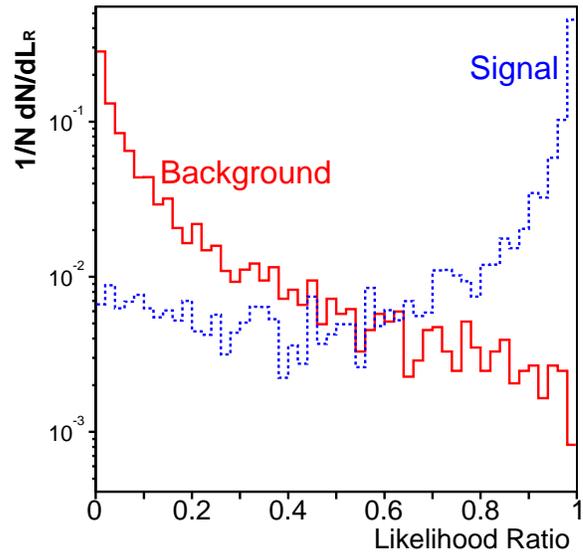}
  \caption{\label{fig:LH} The likelihood ratio distribution for signal (dashed)
  from {\sc pythia} MC and background (solid) from data sidebands in the U-U channel.  
  Similar shapes are observed in the U-X channel.}
\end{figure}

To enhance signal and background separation we construct a multivariate
likelihood ratio based on the input variables: \iso, \pting, and
\ctau\ probability $P(\ctau)=e^{-\ctau/c\tau_{B_{s(d)}}}$, where 
$\tau_{B_{s(d)}}$ is the world average $B_{s(d)}$ lifetime.  We use the $P(\ctau)$ 
variable instead of $\ctau$ in constructing the likelihood ratio because 
the $P(\ctau)$ distribution is nearly flat, and therefore better behaved in
the likelihood.  We define the likelihood ratio to be
\begin{equation}\label{eq:LH}
  L_R = \frac{\prod_{i} \mathbf{P}_{s}(x_{i})}
       {\prod_{i} \mathbf{P}_{s}(x_{i})
      + \prod_{i} \mathbf{P}_{b}(x_{i})},
\end{equation}
where $x_1=\iso$, $x_2=\pting$, $x_3=P(\ctau)$, and $\mathbf{P}_{s(b)}(x_{i})$ 
is the probability that a signal (background) event has an observed $x_i$.  
The probability distributions for the signal events are obtained from the signal MC 
and the background distributions are taken from the data 
sidebands (defined below).  The resulting $L_R$ distributions for the
signal and background events are shown in Fig.~\ref{fig:LH}.

Although a subset of the data was used previously~\cite{PRL1}, we adopt an optimization
strategy that only uses events in the data sidebands in order to avoid biases in
our choice of final selection criteria.
The search window is defined by $5.169 < \Mmm < 5.469\:\GeV/c^2$, corresponding 
to approximately $\pm6$ times the two-track invariant mass resolution, which is 
estimated to be $\sigma_{M}\approx24$~$\mathrm{MeV}/c^{2}$.
The sideband regions $4.669 < \Mmm < 5.169\:\GeV/c^2$ and 
$5.469 < \Mmm < 5.969\:\GeV/c^2$ are used to estimate the background in the
signal region and for extensive analysis cross-checks.  

The \bsmm\ branching fraction is obtained by normalizing to the number of
${\bjk\ra\mm K^+}$ decays collected by the same trigger.  The \bjk\ 
mode is reconstructed using the same baseline requirements as the signal
mode~\cite{bvtx} but including an additional $p_T>1~\GeV/c$ requirement on the kaon  
candidate.  The upper limit on the branching fraction is given by 
\begin{eqnarray}\label{eq:brbsmm}
\brbsmm^{90\% C.L.} = \frac{N_{\bs}^{90\%}}{N_{\bu}}\cdot
     \frac{\alpha_{\bu}}{\alpha_{\bs}}
     \cdot\frac{\epsilon_{\bu}^{base}}{\epsilon_{\bs}^{base}} \cdot 
     \frac{1}{\epsilon_{\bs}^{L_R}} \cdot \nonumber\\
     \frac{f_{u}}{f_{s}}  
     \cdot \mathcal{B}(\bjk), 
\end{eqnarray}
where $N_{\bs}^{90\%}$ is the number of \bsmm\ decays at the  90\% C.L. for  
$N$ observed and $N_{b}$ expected background events. The value of
$N_{\bs}^{90\%}$ is  estimated using 
the Bayesian approach of Ref.~\cite{PDG2004} assuming a flat prior for the 
\brbsmm\ and incorporating Gaussian uncertainties into the limit.  
The number of reconstructed
\bjk\ candidates, $N_{\bu}$, determined from the data using sideband subtraction,
is $1785\pm60$ for the U-U and $696\pm39$ for the U-X channel.  
The parameter $\alpha_{\bs}$ ($\alpha_{\bu}$) is the trigger acceptance and  
$\epsilon_{\bs}^{base}$ ($\epsilon_{\bu}^{base}$) is the efficiency of the
baseline requirements for the signal (normalization) mode.  We apply the 
likelihood ratio requirement only to the signal mode and therefore the 
efficiency of the likelihood ratio ${\epsilon_{\bs}^{L_R}}$ appears only 
in the denominator of Eq.~\ref{eq:brbsmm}.
We use the fragmentation ratio ${f_{u}}/{f_{s}}=3.71\pm0.41$~\cite{FRAG}.  
The branching fraction $\mathcal{B}(\bjk\ra\mm K^{+})=(5.88\pm0.26)\times 10^{-5}$ 
is obtained from Ref.~\cite{PDG2004}.
The expression for \brbdmm\ is derived by replacing \bs\ with \bd\ 
and the fragmentation ratio with ${f_{u}}/{f_{d}}=1$. 

The expected number of background events is estimated by extrapolating 
the number of sideband events passing the baseline requirements to the
signal window and then scaling by the expected rejection factor $\kappa$ for
a given $L_R$ requirement.  The parameter $\kappa$ is determined from the
background $L_R$ distribution, which we generate by randomly sampling the 
\ctau, \pting, and \iso\ distributions from the data sidebands to improve 
statistical precision on $\kappa$.
The relative uncertainty on $N_{b}$ is $\pm15\%$ ($\pm19\%$) for the U-U (U-X)
channel.  The dominant contribution comes from the limited statistics of the
input distributions used to generate the background $L_R$ distribution.

We have cross-checked our background estimate procedure using control samples
from the data: like sign $\mu^{\pm}\mu^{\pm}$ events, \mm\ events with $\ctau<0$, and
a fake-muon enhanced \mm\ sample in which we demand at least one muon 
candidate fail the muon quality requirements.   We compare our background 
predictions to the number of events observed in the search window for a 
wide range of $L_R$ requirements.   No statistically significant 
discrepancies are observed.  For example, requiring $L_R>$0.99 and summing over
all control samples, we predict 3.8 (4.0) background events and observe
4 (3) in the U-U (U-X) channel.
In addition we have considered background contributions from 
$B^{0}_{s,d}\ra\hh$, where $h^{\pm}=\pi^{\pm}$ or $K^{\pm}$, in the 
signal window and determined that the contribution from those decays is 
negligible.

The acceptance ratio ${\alpha_{\bu}}/{\alpha_{\bsd}}$ obtained from
the signal MC is $0.297\pm0.020$ ($0.191\pm0.013$) for the U-U (U-X) 
channel.  The uncertainty includes contributions from systematic variations
of the modeling of the $B$-hadron $p_{T}$ spectrum and longitudinal 
beam profile, and from the statistics of the MC sample. 

The quantity $\epsilon^{base}$ includes the trigger and offline reconstruction 
efficiencies. The trigger efficiency is determined from inclusive data samples
unbiased with respect to the triggers used here.   The ratio of $B^+$ to $B_s^0$ 
trigger efficiencies is measured to be less than 0.1\% away from unity.
We evaluate the single track COT, SVX II, and muon efficiencies using
a data sample of inclusive $J/\psi\ra\mm$ decays collected on single-muon triggers.
The relevant double-track efficiencies are computed by convolution with 
\bsmm\ and \bjk\ MC events surviving the trigger requirements.   The offline 
reconstruction efficiency between signal and normalization mode also largely 
cancels in the ratio with the exception of the kaon efficiency from the \bu\ decay.
Lastly, we obtain the efficiency of the remaining baseline requirements from 
the signal MC and cross-check the results by comparing \bu\ data and MC.  Combining 
all effects, we find ${\epsilon_{\bu}^{base}}/{\epsilon_{\bsd}^{base}}=0.920\pm0.034$ 
($0.915\pm0.034$) for the U-U (U-X) channel.  The uncertainty 
is dominated by systematic uncertainties accounting for kinematic differences
between $J/\psi\ra\mm$ and \bsdmm\ decays.

The efficiency of the likelihood requirement $\epsilon_{\bsd}^{L_R}$ is
estimated from the signal MC.  The efficiency varies from about 70\% for $L_R>0.90$ 
to 35\% for $L_R>0.99$.  We assign a relative systematic uncertainty of $\pm5\%$ 
to both U-U and U-X channels based on comparisons of \bjk\ MC and data samples.

\begin{figure}[h]
  \includegraphics[width=8.0cm]{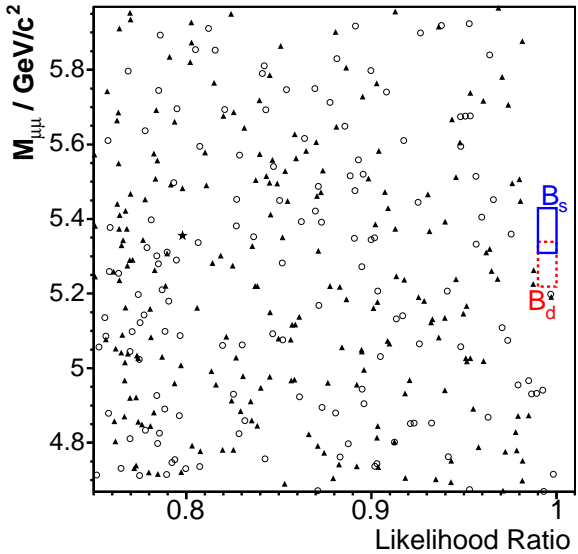}
  \caption{\label{fig:results} The \mm\ invariant mass distribution versus
  likelihood ratio for events satisfying baseline requirements for
  U-U (solid triangle) and U-X (open circle) channels.  
  The \bs\ (solid box) and \bd\ (dashed box) signal regions are also shown.
  The one candidate event observed in the previous analysis is highlighted with 
  a star symbol (at $L_R\approx$0.8).}
\end{figure}

We optimize the analysis based on the {\it{a priori}} expected $90\%$~C.L. 
upper limit on \brbsdmm.  The expected limit for a given set of optimization 
requirements is computed by summing the $90\%$~C.L. limits over all possible 
observations $N$, weighted by the corresponding Poisson probability when 
expecting $N_{b}$.  We scan over a range of $L_R$ requirements and determine 
the optimal value to be $L_R>0.99$.  With the optimized selection requirements, 
the expected number of background events is the same in the \bs\ and \bd\ 
search windows.  Within the signal region of $\pm 60~\MeV/c^2$ 
($\pm 2.5\sigma_{M}$) about the world average \bs\ or \bd\ mass~\cite{PDG2004}, 
$N_{b}$ is $0.81\pm0.12$ [$0.66\pm0.13$], and the
\bs\ single-event-sensitivity is $(1.0\pm0.2) \times 10^{-7}$ 
[$(1.5\pm0.3) \times 10^{-7}$] for the U-U [U-X] channel.  Using 
these criteria we observe no events inside either the \bs\ or \bd\ signal box 
as shown in Fig.~\ref{fig:results}. The one candidate event that survived in the
previous analysis~\cite{PRL1} has an $L_R$ value of $\approx$ 0.8, and 
fail the selection requirements here. 
Using Eq.~\ref{eq:brbsmm} and combining the U-U and U-X channels taking into
account the correlated uncertainties, we derive 90\% (95\%) C.L. limits of
$\brbsmm < 1.5\times 10^{-7}$ $(2.0\times 10^{-7})$ and
$\brbdmm < 3.9\times 10^{-8}$ $(5.1\times 10^{-8})$.  
The new limits improve the previous limits~\cite{D0prl,BABARprl} by a factor 
of two and can be used to reduce the allowed parameter 
space of a broad spectrum of SUSY models~\cite{SO10,RPV,Models}.

We thank the Fermilab staff and the technical staffs of the participating 
institutions for their vital contributions. This work was supported by the U.S. 
Department of Energy and National Science Foundation; the Italian Istituto 
Nazionale di Fisica Nucleare; the Ministry of Education, Culture, Sports, 
Science and Technology of Japan; the Natural Sciences and Engineering Research 
Council of Canada; the National Science Council of the Republic of China; the 
Swiss National Science Foundation; the A.P. Sloan Foundation; the 
Bundesministerium f\"ur Bildung und Forschung, Germany; the Korean Science 
and Engineering Foundation and the Korean Research Foundation; the Particle 
Physics and Astronomy Research Council and the Royal Society, UK; the 
Russian Foundation for Basic Research; the Comisi\'on Interministerial de 
Ciencia y Tecnolog\'{\i}a, Spain; in part by the European Community's Human 
Potential Programme under contract HPRN-CT-2002-00292; and the Academy of Finland. 



\begin{thebibliography}{99.}

\bibitem{qconj}
 Throughout this paper inclusion of charge conjugate modes is implicit.

\bibitem{smbr}
  G.~Buchalla and A.~J.~Buras,  Nucl.\ Phys.\ {\bf B400}, 225 (1993);
  A.J.~Buras, Phys.\ Lett.\ B {\bf{566}}, 115 (2003).

\bibitem{D0prl}
 D0 Collaboration, V.~Abazov {\it{et al.}}, 
 Phys.\ Rev.\ Lett.\ {\bf{94}}, 071802 (2005).

\bibitem{BABARprl}
 BABAR Collaboration, B.~Aubert {\it{et al.}},
 Phys.\ Rev.\ Lett {\bf{94}}, 221803 (2005).

\bibitem{tanb}
 S.~Choudhury and N.~Gaur, Phys.\ Lett.\ B {\bf{451}}, 86 (1999);
 K.S.~Babu and C.~Kolda, Phys.\ Rev.\ Lett.\ {\bf{84}}, 228 (2000).

\bibitem{SO10}
 R.~Dermisek {\it{et al.}}, J.\ High\ Energy\ Phys.\ {\bf{04}}, 037 (2003);
 D.~Auto {\it{et al.}}, J.\ High\ Energy\ Phys.\  {\bf{06}}, 023 (2003);
 T.~Blazek {\it{et al.}}, Phys.\ Lett.\ B {\bf{589}}, 39 (2004).
 
\bibitem{RPV}
 R.~Arnowitt {\it{et al.}}, Phys.\ Lett.\ B {\bf{538}}, 121 (2002).

\bibitem{Models}
 H.~Baer {\it{et al.}}, J.\ High\ Energy\ Phys.\  {\bf{07}}, 050 (2002);
 S.~Baek, P.~Ko, and W.Y.~Song, Phys.\ Rev.\ Lett.\ {\bf{89}}, 271801 (2002);
 H.~Logan and U.~Nierste, Phys. Lett.\ B {\bf{566}}, 115 (2003);
 A.~Dedes and B.~Huffman, Phys.\ Lett.\ B {\bf{600}}, 261 (2004);
 G.L.~Kane, C.~Kolda, and J.E.~Lennon, hep-ph/0310042.

\bibitem{PRL1}
  CDF~Collaboration, D.~Acosta {\it{et al.}}, 
  Phys.\ Rev.\ Lett.\ {\bf{93}}, 032001 (2004).

\bibitem{Acosta:2004yw}
  CDF~Collaboration, D.~Acosta {\it et al.},
  Phys.\ Rev.\ D {\bf 71}, 032001 (2005).

\bibitem{SVX}
 A.~Sill {\it{et al.}}, Nucl.\ Instrum.\ Meth. A {\bf{447}}, 1 (2000).

\bibitem{COT}
 T.~Affolder {\it{et al.}}, Nucl.\ Instrum.\ Meth.\ A {\bf{526}}, 249 (2004).

\bibitem{CMU}
 G.~Ascoli {\it{et al.}}, Nucl.\ Instrum.\ Meth. A {\bf{268}}, 33 (1988).

\bibitem{pythia}
 T.~Sj\"{o}strand {\it{et al.}}, Comp.\ Phys.\ Commun.\ {\bf{135}}, 238 (2001).

\bibitem{iso}
 The $B$-candidate isolation is defined as 
 $\iso=| \ptmm | /(\sum_{i}p^{i}_{T} + | \ptmm | )$, where the
 sum is over all tracks with $\sqrt{\Delta\eta^2+\Delta\phi^2} \le 1$; 
 $\Delta\phi$ and $\Delta\eta$ are the azimuthal angle and
 pseudorapidity of track $i$ with respect to \pmm.
 Also see V.~Krutelyov, Ph.D. Thesis, Texas A\&M University, 2005 (unpublished).

\bibitem{bvtx}
  We only use the two muon tracks in the $B^+$ vertex fit.

\bibitem{PDG2004}
  S.~Eidelman {\it et al.},
  Phys.\ Lett.\ B {\bf 592}, 1 (2004).


\bibitem{FRAG}
 $f_x$ is the fraction of weakly decaying $B_x$ hadron in $b$ quark 
 fragmentation. We use values from~\cite{PDG2004} and take into account the 
 correlations between $f_s$ and $f_u$.

\end{thebibliography}

\end{document}